\documentclass[prl,superscriptaddress,twocolumn]{revtex4}
\pdfoutput=1
\usepackage{amsfonts}
\usepackage{graphicx}
\usepackage{amsmath}
\usepackage{xcolor}
\usepackage{bm}

\DeclareMathOperator{\tr}{tr}
\DeclareMathOperator{\Ca}{{\rm Ca}}

\begin{document}

\title{Spontaneous Division and Motility in Active Nematic Droplets}

\author{Luca Giomi}
\affiliation{SISSA, International School for Advanced Studies, Via Bonomea 265, 34136 Trieste, Italy}
\author{Antonio DeSimone}
\affiliation{SISSA, International School for Advanced Studies, Via Bonomea 265, 34136 Trieste, Italy}

\begin{abstract}
We investigate the mechanics of an active droplet endowed with internal nematic order and surrounded by	an isotropic Newtonian fluid. Using numerical simulations we demonstrate that, due to the interplay between the active stresses and the defective geometry of the nematic director, this system exhibits two of the fundamental functions of living cells: spontaneous division and motility, by means of self-generated hydrodynamics flows. These behaviors can be selectively activated by controlling a single physical parameter, namely, an active variant of the capillary number.
\end{abstract}

\maketitle
The goal of understanding the machinery of life has led in recent years to the ambitious idea of constructing synthetic \emph{minimal cells}: chemical machines capable of reproducing some of the fundamental traits of living cells such as self-maintaining, duplicating, and passing information across generations \cite{Szostak:2001,Hutchison:1999}. Whether the ultimate goal is to design engineered cells for medical applications or to reconstruct the history of life from its prebiotic origin to its present complexity, understanding the architecture and operation of minimal cellular machines represents a fundamental step towards unraveling the transition from inanimate to living matter \cite{Rasmussen:2003}. 

The first challenge in this program is to identify the crossover region between molecular self-assembly and molecular operation \cite{Fellermann:2007}, a task that has led to investigate the design principles and the mechanics of self-dividing lipid vesicles \cite{Hanczyc:2004,Takakura:2004} and more recently the coupling of self-dividing vesicles with self-replicating nucleic acids enclosed in the interior of the vesicle \cite{Kurihara:2011,Luisi:2011}. Mass transfer due to hydrodynamic instabilities, such as the Marangoni effect, has been recently invoked as a possible route to motility in prebiotic structures \cite{Hanczyc:2011}. 

In this article, we would like to propose the mechanics of active liquid crystals
as a possible framework to study the transition between nonliving and living matter, possibly providing some principles for the design of artificial cells. Active liquid crystals are non-equilibrium fluids composed of orientationally ordered self-driven particles. Each active particle is capable of converting stored or ambient energy into systematic movement. The interaction of active particles with each other and with the surrounding medium gives rise to mechanical stresses and highly correlated collective motion over  large scales \cite{Kruse:2004,Giomi:2011,Woodhouse:2012,Forest:2013}. Originally developed for modeling collections of swimming \cite{Pedley:1992} and crawling cells \cite{Gruler:1999}, and later extended to the cytoskeleton and its components \cite{Lee:2001,Liverpool:2003,Kruse:2004}, the mechanics of active liquid crystals has gained increasing attention in the last decade thanks to its successes in the modeling of cellular motility, intracellular movement and transport \cite{Marchetti:2013}. Thus, a powerful tool for both understanding the behavior of biological systems and designing novel, biologically inspired, materials has now become available.

Here we consider a two-dimensional active droplet endowed with internal nematic order and surrounded by an isotropic Newtonian fluid.  We demonstrate that, due to the interplay between the active stresses and the geometry of the nematic director, which is constrained by the droplet topology, this system exhibits two of the defining functions of living cells: spontaneous division and motility, by means of self-generated hydrodynamic flows. These behaviors can be selectively activated by controlling a single physical parameter corresponding to an active variant of the {\em capillary number}.
%starting from  a quiescent state in which the droplet does not divide nor move. 
While the task of implementing this strategy in synthetic biology is  far from trivial, the fact that two of the fundamental cellular processes can arise from a single physical mechanism is remarkable and can possibly provide new insight on the transition from prebiotic structures to living micro-organisms.
 
The hydrodynamic equations of an active nematic medium have been proposed based on phenomenological arguments \cite{Kruse:2004,Marenduzzo:2007,Giomi:2011,Giomi:2012}, or derived from microscopic models \cite{Baskaran:2008,Lau:2009,Bertin:2013}. Apart from some unimportant differences, these models agree in identifying  the so called ``active stress'' (proportional to the local nematic order) as the fundamental non-equilibrium contribution due to molecular activity. The  degrees of freedom needed to describe an active nematic fluid are then the flow velocity field $\bm{v}$ and  the nematic tensor field $\bm{Q}$ which, for uniaxial nematics in two dimensions, is given by $Q_{ij} = S\left(n_{i}n_{j}-\delta_{ij}/2\right)$, where $\bm{n}$ is the nematic director and $0 \le S \le 1$ is the order parameter representing the local extent of nematic order. Here we consider the case of an incompressible two-phase fluid consisting of a nematic phase embedded in an isotropic phase. The two phases have, for simplicity, the same density $\rho$, which is then constant throughout the system. 

In order to implement the mechanism of phase separation we use a diffuse interface method similar to that proposed in Ref. \cite{Yue:2004} to simulate two-phase flows in complex fluids. In this picture the two phases are described by a phase field variable $-1 \le \phi \le 1$ such that $\phi=-1$ represents the isotropic phase, $\phi=1$ the nematic phase and $\phi\approx 0$ the diffuse interface. The effective capillarity of the interface can be described starting from a Ginzburg-Landau energy density of the form
\begin{equation}\label{eq:energy_capillarity}
f_{\rm cap} = \frac{1}{2}\kappa\left[|\nabla\phi|^{2}+\frac{1}{2\epsilon^{2}}\,(\phi^{2}-1)^{2}\right]\;.
\end{equation}
This functional favors the separation of the phases into domains of pure components (i.e. $\phi=\pm 1$). The classic surface tension $\Sigma$ of the interface is related to the parameters appearing in Eq. \eqref{eq:energy_capillarity} by $\Sigma=\sqrt{8}/3\,(\kappa/\epsilon)$ \cite{Modica:1987,Yue:2004,Yang:2011}. Furthermore, the incompressibility of the fluid phases implies $\int dA\,\phi = {\rm constant}$. The interfacial tension contributes to the total mechanical stress with a term $\bm{\sigma}^{\rm c}$. This can be found by equating the energy variation under a small virtual displacement to the work of the restoring forces \cite{Anderson:1998}. This yields the following expression for the body force due to capillarity $\bm{f} = \nabla\cdot\bm{\sigma}^{\rm c} = -\phi\nabla\mu$ where $\mu = \delta F_{\rm cap}/\delta \phi = - \kappa \left[\Delta \phi-\phi(\phi^{2}-1)/\epsilon^{2}\right]$ is an effective chemical potential. This force is experienced by the system only along the diffuse interface where $\mu$ undergoes an abrupt spatial variation. The hydrodynamic equations for the fields $\phi$, $\bm{Q},$ and the flow velocity $\bm{v}$ are then given by \cite{Giomi:2011,Giomi:2012}:
\begin{subequations}\label{eq:hydrodynamics}
\begin{gather}
\frac{D\phi}{Dt} = M \kappa \left[\Delta \phi-\frac{\phi(\phi^{2}-1)}{\epsilon^{2}}+\xi(\phi)\right]\;,\\
\rho \frac{Dv_{i}}{Dt}=\eta\Delta v_{i}-\partial_{i}p-\phi\partial_{i}\mu+\partial_{j}\sigma_{ij}\;,\\
\frac{D{Q}_{ij}}{Dt}=\lambda S u_{ij}+Q_{ik}\omega_{kj}-\omega_{ik}Q_{kj}+\gamma^{-1}H_{ij}\;,
\end{gather}
\end{subequations}
where $D/Dt=\partial_{t}+\bm{v}\cdot\nabla$ is the material time derivative, $M$ is a mobility coefficient, $\eta$ the viscosity (also assumed to be the same in both fluids) and $p$ the pressure. The function $\xi$ is a Lagrange multiplier that guarantees mass conservation:
\begin{equation}\label{eq:lagrange}
\xi(\phi) = |\phi^{2}-1|\,\frac{\int dA\,\phi(\phi^{2}-1)}{\int dA\,|\phi^{2}-1|}\;.
\end{equation}
This form was recently introduced in Ref. \cite{Brassel:2011} as an alternative to the more classic non-local expression, $\int dA\,\phi(\phi^{2}-1)/\int dA$ \cite{Rubinstein:1992}.  Eq. \eqref{eq:lagrange}  leads to  higher accuracy in mass conservation by combining both local and non-local terms. In addition, we noticed that Eq. \eqref{eq:lagrange} allows us to simulate smaller droplets compared its non-local counterpart by hindering the stability of the trivial solution where $\phi$ is uniform and the two phases mix. In Eq. (\ref{eq:hydrodynamics}c) $u_{ij}=(\partial_{i}v_{j}+\partial_{j}v_{i})/2$ and  $\omega_{ij}=(\partial_{i}v_{j}-\partial_{j}v_{i})/2$ are the rate of strain  and the vorticity tensors, respectively, and represent the coupling between orientational order and flow (with $\lambda$ the flow alignment parameter \cite{DeGennes:1993}). The molecular field $H_{ij}$, on the other hand, drives the relaxational dynamics of the nematic phase (with $\gamma$ a rotational viscosity) and can be obtained from the variation of the total free energy of the nematic phase $F_{\rm nem} = \int dA\,(f_{\rm LdG}+f_{\rm anc})$ as $H_{ij}=-\delta F_{\rm nem}/\delta Q_{ij}$. Here the Landau-de Gennes free energy density $f_{\rm LdG}$ governs the behavior of the bulk nematic phase:
\begin{equation}\label{eq:energy_ldg}
f_{\rm LdG}=\frac{1}{2}K\left[|\nabla\bm{Q}|^{2}+\frac{1}{\delta^{2}}\tr\bm{Q}^{2}\left(\tr\bm{Q}^{2}-\phi\right)\right]\;,
\end{equation}
where $K$ is an elastic constant (proportional to the classic Frank constant) and the second terms in Eq. \eqref{eq:energy_ldg} leads to a second order isotropic/nematic phase transition controlled by the phase-field $\phi$. Since $\tr\bm{Q}^{2}=S^{2}/2$, Eq. \eqref{eq:energy_ldg} implies that, where $\phi=-1$, $f_{\rm LdG}$ has a minimum for $S=0$, corresponding to the isotropic phase, and for $\phi=1$, $f_{\rm LdG}$ is minimized by $S=\sqrt{\phi}=1$. 

The term $f_{\rm anc}$ represents the anchoring energy at the isotropic/nematic interface. Here we use a diffuse version of the Nobili-Durand anchoring energy \cite{Nobili:1992}:
\begin{equation}\label{eq:energy_anchoring}
f_{\rm anc} = \frac{1}{2}W\tr(|\nabla\phi|^{2}\bm{Q}-\bm{A})^{2}	
\end{equation}
where $A_{ij}=\partial_{i}\phi\,\partial_{j}\phi-|\nabla\phi|^{2}\delta_{ij}/2$. The effect of
$f_{\rm anc}$ is to favor a director field $\bm{n}$  parallel to  $\nabla\phi$ (hence normal to the interface) and the value  $S=1$ for the nematic order parameter.

\begin{figure}[t]
\centering
\includegraphics[width=0.8\columnwidth]{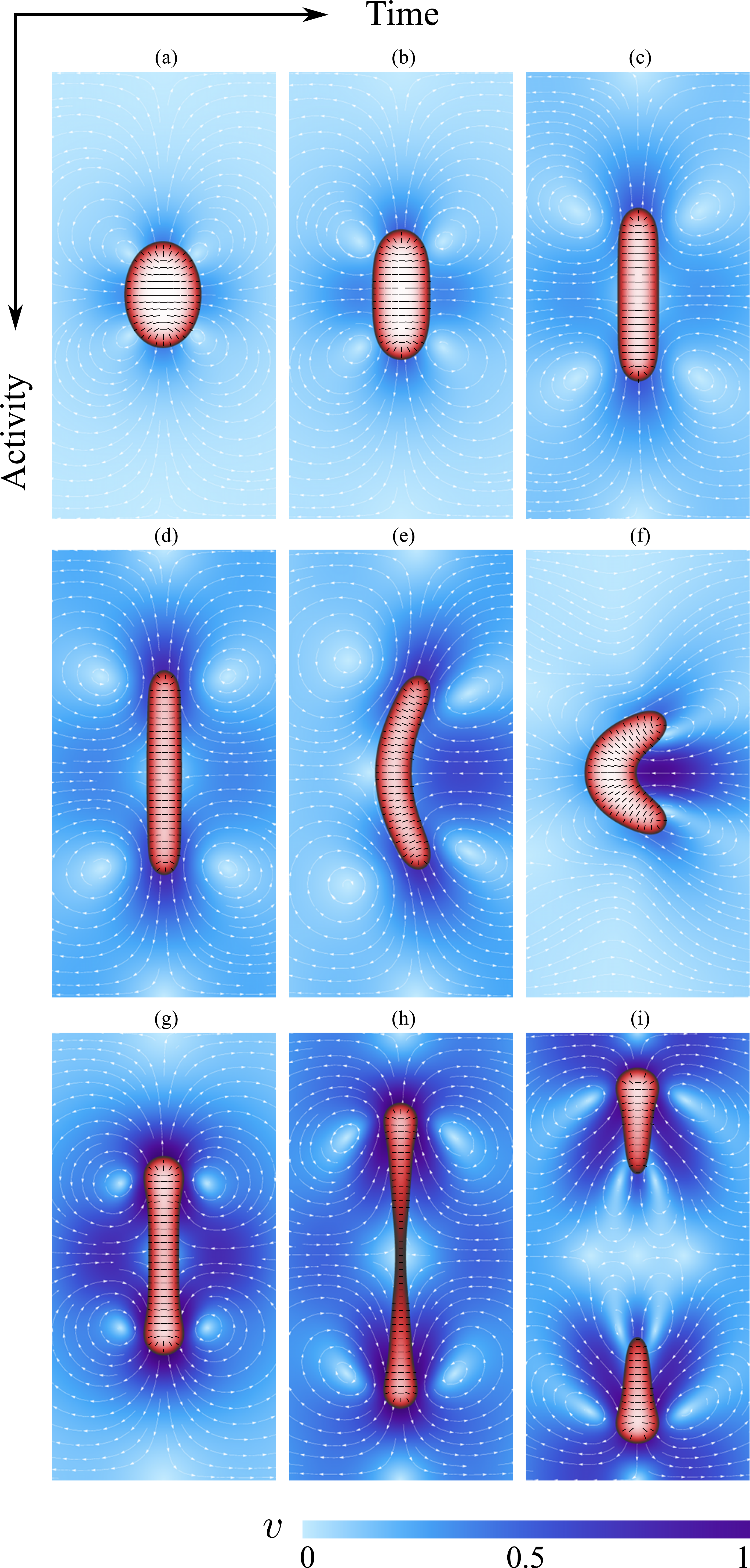}
\caption{\label{fig:droplet}The three behaviors of an active nematic droplet for fixed surface tension $\Sigma=2.6$ and varying activity obtained from a numerical solution of Eq. \eqref{eq:hydrodynamics}. (a-c) For small activity the droplet stretches under the effect of the quadrupolar straining flow generated by the pair of $+1/2$ disclinations. (d-f) For $\alpha=16$, the uniformly oriented director field in the interior of the droplet is unstable to splay and the droplet deforms. Following the deformation of the droplet, the backflow is no longer axially symmetry and this causes the droplet to move. (g-i) For very large activity ($\alpha=36$), the capillary forces are no longer sufficient to balance the initial straining flow and the droplet divides.}
\end{figure}

Finally, the stress tensor $\bm{\sigma}=\bm{\sigma}^{c}+\bm{\sigma}^{\rm r}+\bm{\sigma}^{\rm a}$ is the sum of the capillary stress, the elastic stress due to nematic elasticity $\sigma^{\rm r}_{ij}=-\lambda S H_{ij}+Q_{ik}H_{kj}-H_{ik}Q_{kj}$, and  of an active contribution $\sigma^{\rm a}_{ij}=\alpha\,Q_{ij}$ that describes contractile ($\alpha>0$) and tensile ($\alpha<0$) stresses exerted by the active particles in the direction of the director field. 

We have integrated Eqs.~\eqref{eq:hydrodynamics} numerically in a square $L\times L$ domain with periodic boundary conditions. The initial configuration consists of a circular droplet of radius $R=L/10$, with director field  uniformly aligned along the $x-$axis,  and the flow velocity identically zero. The integration is performed by using the finite difference scheme described in Refs.~\cite{Giomi:2011,Giomi:2012}. To make Eqs.~\eqref{eq:hydrodynamics} dimensionless, we normalize distance by $R$, time by $\tau=\gamma R^{2}/K$ corresponding to the relaxation time scale of the nematic phase over the length scale of the droplet, and stress by the elastic stress $\sigma=K/R^{2}$. All the other quantities are rescaled accordingly. 
 
We have focused on the interplay between the surface tension $\Sigma$ of the droplet and the active stress $\alpha$ and kept the other parameters constant ($\lambda=0.1$, $\eta=M=1$, $W=1.25$, $\epsilon=\delta=0.15$). It is well known that, in bulk systems, contractile and tensile active stresses favor respectively splayed and bent configurations of the nematic director through feedback mechanisms mediated by the flow \cite{Voituriez:2005}. As a consequence, a uniformly oriented reference configuration becomes unstable once the internal active stress exceeds a critical value $\alpha_{\rm c} \sim \eta / \tau$ \cite{Giomi:2011,Giomi:2012}. 

In nematic droplets, however, the homeotropic anchoring changes this picture significantly because it promotes the formation of topological defects. As a consequence of the disk topology and the normal orientation at the interface, the director field in the interior of the droplet is forced to form a defective texture whose total winding number is $k_{\rm tot}=1/(2\pi)\oint d\theta = 1$, where $\theta$ is the angle between $\bm{n}$ and an arbitrary axis and the integral is calculated along the droplet boundary. This is achieved by forming two $+1/2$ disclinations approximately located at a distance of order  $\epsilon=\delta\sqrt{W/K}$ from the droplet boundary (see Fig. \ref{fig:droplet}a). In passive nematic droplets, the defects repel each other with a force inversely proportional to their distance. This repulsion is in turn balanced by surface tension leading to a slight elongation of the droplet along the line joining the defects \cite{Yue:2004}. 

The scenario outlined above is dramatically altered by the presence of activity. Fueled by the strong distortion introduced by a defect, the active stresses give rise to a flow whose magnitude and direction is controlled by the activity constant $\alpha$ \cite{Giomi:2013}. For a contractile droplet ($\alpha>0$) with homeotropic boundary, the axisymmetric structure of the director drives a typical quadrupolar straining flow, causing a much more drastic elongation than that produced by the elastic repulsion alone (Fig. \ref{fig:droplet}c). 

\begin{figure}[t]
\centering
\includegraphics[width=.8\columnwidth]{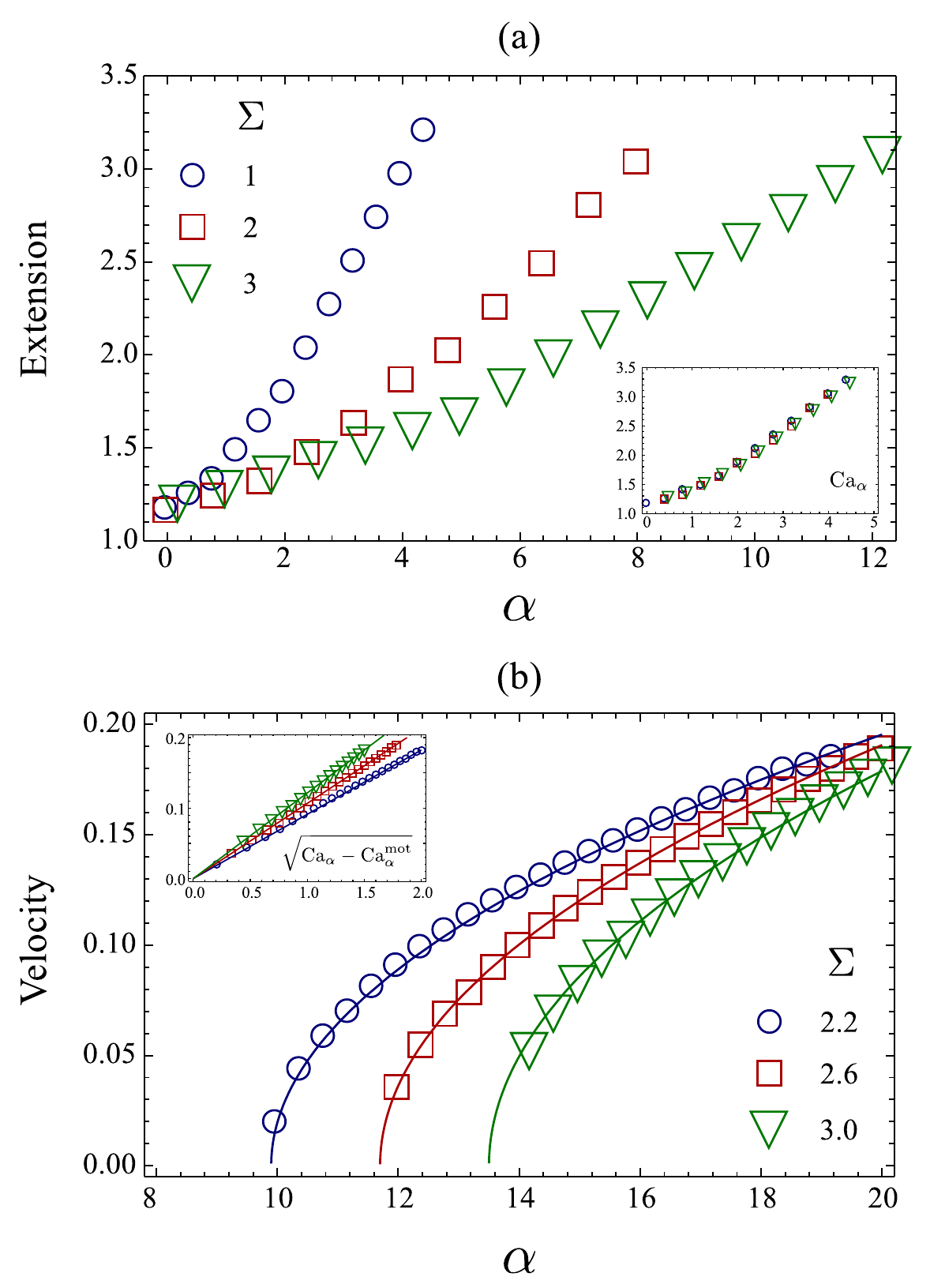}
\caption{\label{fig:force_velocity}(a) Extension of the droplet (measured as the distance between the defects) versus activity for various $\Sigma$ values. The data collapse on the same master curve when rescaled with respect to the active capillary number ${\rm Ca}_{\alpha}=\alpha R/\Sigma$ (inset). (b) The velocity of a motile droplet versus activity for various $\Sigma$ value. When rescaled with respect to ${\rm Ca}_{\alpha}$ the data intersect at the critical capillary number ${\rm Ca}_{\alpha}^{\rm mot}\approx 4.5$ (inset). The solid lines shows the typical square-root law and are obtained from a fit.}
\end{figure}

To characterize the spontaneous deformation we have measured the extension of the droplet as a function of the activity parameter $\alpha$, for various $\Sigma$ values (Fig. \ref{fig:force_velocity}a). This shows a clear linear behavior except for small $\alpha$ values, where the deformation is mainly dictated by the elastic repulsion between the defects. This behavior is consistent with the general picture of drop deformation in a straining flow \cite{Rallison:1984,Stone:1994}. According to this, a neutrally buoyant droplet placed in a shear flow experiences a strain that scales linearly with the capillary number $\Ca =\eta v/\Sigma$, where $v$ is the typical flow velocity. Now, the velocity of the flow generated by an active nematic disclination scales like $v \sim \alpha R/\eta$ \cite{Giomi:2013}, hence the linear dependence of the droplet extension on $\alpha$. Moreover, by introducing an active variant of the capillary number, defined as $\Ca_{\alpha}=\alpha R/\Sigma$, one can rescale the numerical data and collapse them on the same master curve (Fig. \ref{fig:force_velocity}a, inset). This indicates that the {\rm active capillary number} ${\rm Ca}_{\alpha}$, expressing the ratio between active and capillary forces, is the fundamental degree of freedom of active droplets in this stationary regime.

For larger activity the droplet becomes motile. Like in the case of active polar droplets \cite{Tjhung:2012,Whitfield:2013}, motility is achieved by means of a spontaneous splay deformation arising from the instability of the configuration of lowest nematic energy  (for which $H_{ij}=0$). As for static deformations, the droplet initially elongates as a consequence of the quadrupolar straining flow driven by the defects (Fig. \ref{fig:droplet}d). In the configuration of maximal elongation, the director field in the interior of the droplet is rather uniform, but after some time this uniform configuration starts to spontaneously splay (Fig. \ref{fig:droplet}e). The splayed configuration of the director field breaks the axial symmetry of the systems and transforms the quadrupolar flow in a dipolar flow consisting of two large vortices running across the droplet and tilted in such a way to form a V-pattern that follows the shape of the droplet. This causes the droplet to move at constant velocity along the axis of the V, where the two vortices meet and the flow velocity is maximal, in the direction of the tip. Fig. \ref{fig:force_velocity}b shows a plot of the center of mass velocity versus $\alpha$ for various $\Sigma$ values; due to the initial axial symmetry, the onset of motion occurs as a supercritical bifurcation and the velocity follows a typical square-root scaling law. Unlike the extension curve, the velocity data do not collapse on the same curve when rescaled with respect to $\Ca_{\alpha}$, but they do intersect at the critical point (Fig. \ref{fig:force_velocity}b, inset). This implies that, while activity and surface tension independently affect the motion of active droplets, the onset of motility is controlled uniquely by the active capillary number $\Ca_{\alpha}$. With the choice of parameters used here the motility transition occurs at $\Ca_{\alpha}^{\rm mot}\approx 4.5$. 

\begin{figure}[t]
\centering
\includegraphics[width=0.7\columnwidth]{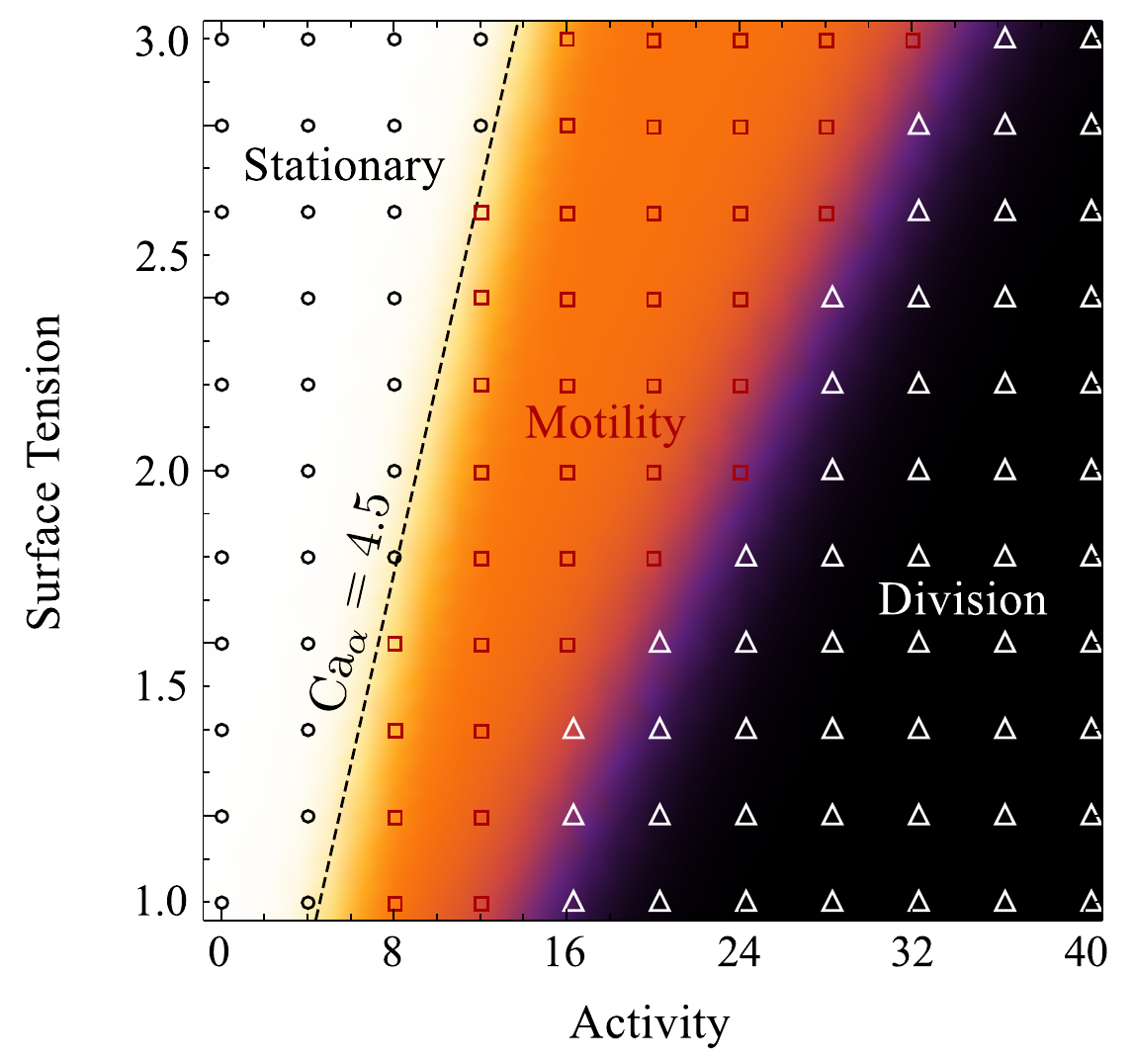}
\caption{\label{fig:phase_diagram}Phase diagram showing the three classes of behavior exhibited by contractile active droplets for different  $\alpha$ (activity) and $\Sigma$ (surface tension) values obtained by numerically solving Eqs. \eqref{eq:hydrodynamics}. For low activity, the quadrupolar straining flow generated by the pair of $+1/2$ disclinations leads to a stationary elongated shape (white region, black circles). When the activity is very strong, the active backflow causes the droplet to spontaneously divide (black region, white triangles). According to the strength of activity the division can produce two or more daughter droplets. For intermediate activity and sufficiently large surface tension the director spontaneously splays and the droplet moves as a consequence of the associated backflow.}
\end{figure}

Motility occurs as the combination of two processes: the initial elongation of the droplet, driven by the straining flow produced by the defects, and the instability of this configuration to splay. The existence of the intermediate elongated configuration is guaranteed by the fact that viscous and pressure forces exerted by the flow on the droplet are balanced by the resistance due to interfacial tension forces. For large capillary numbers, the capillary forces are no longer sufficient to achieve this balance, the droplet continuously stretches and eventually divides {\em before} the splay instability can develop (Fig. \ref{fig:droplet}g-i). Fig. \ref{fig:phase_diagram} shows a phase diagram in the $(\alpha,\Sigma)-$plane summarizing the three behaviors described so far. The approximately straight phase boundary between the motility and division regimes, suggests that, like for the onset of motility, division might also be controlled solely by $\Ca_{\alpha}$. Our numerical data indicate that division occurs at a critical value $11.4 < \Ca_{\alpha}^{\rm div} < 16$. Once the parent droplet first divides, the active capillary number drops due to the reduction in the droplet size $R$. Thus the two daughter droplets remain stable unless the activity is large enough that the new capillary number is itself larger than $\Ca_{\alpha}^{\rm div}$, in which case multiple divisions occur. This mechanism can in principle lead to a cascade of divisions that terminates only once the size of the youngest generation of droplets is such that $\Ca<\Ca_{\alpha}^{\rm div}$. A detailed analysis of this phenomenon is however outside the scope of this work. We note that the existence of spontaneous division in active nematic droplets has the further consequence of setting a limit speed for activity-driven motility as division replaces motility at large enough activities.

In summary, we have investigated the mechanics of a contractile active nematic droplet surrounded by a Newtonian fluid. Due to the interplay between the active stresses and the defective geometry of the nematic director, the system exhibits two of the defining functions of living cells: spontaneous division and motility, which can be selectively activated by controlling a single physical parameter: the active capillary number. While the physical mechanisms involved in these processes are far from trivial, the fact that such a remarkable cell-like behavior can be achieved autonomously and without a central control could shed light on the transition from inanimate to living matter and possibly inspire new principles for the design of artificial cells.

LG is grateful to Rastko Sknepnek and Luca Heltai for help with the simulations. This article is dedicated to the memory of Giovanni Romeo, whose unforgettable passion and curiosity have been inspirational throughout this work.

\end{document}